\documentclass[%
 reprint,
 amsmath,amssymb,
 prl,
]{revtex4-1}

\usepackage{graphicx}
\usepackage{dcolumn}
\usepackage{bm}
\usepackage{amssymb}
\usepackage{epstopdf}

\begin{document}


\title{A Unified Framework for Information Coding:\\
Oscillations, Memory, and Zombie Modes}

\author{Andrew T Sornborger}
\email{ats@math.ucdavis.edu}
\affiliation{Department of Mathematics, University of California, Davis, USA}%


\author{Louis Tao}
\email{taolt@mail.cbi.pku.cn.edu}
\affiliation{Center for Bioinformatics, National Laboratory of Protein Engineering and Plant Genetic Engineering, College of Life Sciences, and Center for Quantitative Biology, Peking University, Beijing, China}%

\date{\today}

\begin{abstract}
\noindent
Synchronous neural activity can improve neural processing and is believed to mediate neuronal interaction by providing temporal windows during which information is more easily transferred. We demonstrate a pulse gating mechanism in a feedforward network that can exactly propagate graded information through a multilayer circuit. Based on this mechanism, we present a unified framework wherein neural information coding and processing can be considered as a product of linear maps under the active control of a pulse generator. Distinct control and processing components combine to form the basis for the binding, propagation, and processing of dynamically routed information within neural pathways. Using our framework, we construct example neural circuits to 1) maintain a short-term memory, 2) compute time-windowed Fourier transforms, and 3) perform spatial rotations.  We postulate that such circuits, with stereotyped control and processing of information, are the neural correlates of Crick and Koch's zombie modes.\end{abstract}

\pacs{Valid PACS appear here}
\maketitle


\section{Introduction}

Accumulating experimental evidence implicates coherent activity as an important element of cognition. Since their discovery \citep{GrayEtAl1989}, gamma band oscillations have been demonstrated to exist in hippocampus \citep{BraginEtAl1995,CsicsvariEtAl2003,ColginEtAl2009}, 
visual cortex \citep{GrayEtAl1989,Livingstone1996,WomelsdorfEtAl2007}, 
auditory cortex \citep{BroschEtAl2002}, 
somatosensory cortex \citep{BauerEtAl2006}, 
parietal cortex \citep{PesaranEtAl2002,BuschmanMiller2007,MedendorpEtAl2007}, 
various nodes of the frontal cortex \citep{BuschmanMiller2007,GregorgiouEtAl2009,SohalEtAl2009}, 
amygdala and striatum\citep{PopescuEtAl2009}. Gamma oscillations sharpen orientation \citep{AzouzGray2000} and contrast\citep{HenrieShapley2005} tuning in V1, and speed and direction tuning in MT \citep{LiuNewsome2006}. Attention has been shown to enhance gamma oscillation synchronization in V4, while decreasing low-frequency synchronization \citep{FriesEtAl2001,FriesEtAl2008} and to increase synchronization between V4 and FEF \citep{GregorgiouEtAl2009}, LIP and FEF \citep{BuschmanMiller2007}, V1 and V4\citep{BosmanEtAl2012}, and MT and LIP \citep{SaalmannEtAl2007}; Interactions between sender and receiver neurons are improved when consistent gamma-phase relationships exist between two communicating sites \citep{WomelsdorfEtAl2007}.

Theta-band oscillations have been shown to be associated with visual spatial memory \citep{OKeefe1993,Buzsaki2002}, where neurons encoding the locations of visual stimuli and an animal's own position have been identified \citep{OKeefe1993,SkaggsEtAl1996}. Additionally, loss of theta gives rise to spatial memory deficits \citep{Winson1978} and pharmacologically enhanced theta improves learning and memory \citep{MarkowskaEtAl1995}.

These experimental investigations of population oscillations in distinct brain regions have informed the modern understanding of information coding in neural systems \citep{QuianQuirogaPanzeri2013,pmid20725095}. Theoretical mechanisms have been proposed for short-term memory, information transfer via spike coincidence  and information gating that rely on gamma- and theta-band oscillations. For example, Abeles's synfire network \citep{Abeles1982,KonigEtAl1996,pmid10591212,pmid21106815,KistlerGerstner2002}, the Lisman-Idiart interleaved-memory (IM) model \citep{LismanIdiart1995},  and Fries's communication-through-coherence (CTC) model \citep{Fries2005} all make use of the fact that coherent oscillations can provide windows in time during which spikes may be more easily transferred through a neural circuit. Thus, neurons firing synchronously can transfer their activity quickly downstream. However, the precise mechanism  and the extent to which the brain can make use of coherent activity to transfer information have remained unclear.

Recently, we have shown that, with coordinated gating pulses, graded current amplitudes can be {\it exactly} transferred between neural populations. We demonstrated this mechanism in both mean-field and integrate-and-fire (I\&F) models and showed that it is robust to pulse timing inaccuracies, mis-tuned synaptic strengths and finite size effects. We then proposed an information coding scheme in which information {\it control} is segregated from information {\it content} and {\it processing}. The resulting theoretical framework allows for a general setting in which to consider neural circuits. 

In this paper, we describe a second mechanism capable of exact, graded firing rate transfer. Additionally, we show that a neural circuit in our framework is equivalent to a set of iterated linear maps controlled by one or more pattern generators. Since it is closely related to the IM and CTC models, our framework naturally contains the conceptual constructs believed to be related to neural oscillations, including information gating, encoding and memory. Furthermore, our framework provides an understanding of how functional connectivity may be rapidly switched to dynamically route information in a flexible way.

We emphasize that our framework is not equivalent to a set of concatenated filters because of the active control element. We propose that a circuit that embodies our information coding paradigm can be understood as a {\it zombie mode}, one of the neural correlates of consciousness responsible for cognitive automaticity outlined by Crick and Koch \citep{CrickKoch2003}. Zombie modes can be thought of as cognitive processing reflexes which have been learned and no longer need conscious control.

Via a set of examples, we demonstrate the propagation, processing and dynamical routing capabilities of our framework. We present a graded memory circuit that generalizes proposals such as the IM model and extends other graded memory schemes \citep{SeungEtAl2000,Goldman2008}. We construct a circuit capable of continuously analyzing streaming input. Finally, we construct a reentrant circuit capable of performing general spatial rotations on a coordinate vector.

Subsequent papers will use our framework to 1) construct a zombie mode that can encode streaming visual information and, using predictive coding, decode and bind the input to form a representation of object location, and 2) demonstrate how egocentric coordinates can be transformed into allocentric coordinates using appropriate control mechanisms.

\section{Results}

\subsection{Exact Firing Rate Amplitude Transfer via Pulse Gating}

In a previous paper, we showed that information contained in the amplitude of a neural current may be {\it exactly} transferred from one neuronal population to another, as long as well-timed current pulses are injected into the populations. We derived this pulse-based transfer mechanism using mean-field equations for a current-based integrate-and-fire neuronal network.

We have also discovered a somewhat simpler model based on the transfer of firing rate amplitudes that we present here.

In a pulse-gated feed-forward neuronal network, firing rates for population $j+1$, synaptically coupled to population $j$, obey
$$\tau \frac{d}{dt}m_{j+1} = -m_{j+1} + S f\left(p_{j+1}(t) + m_{j}\right) \; ,$$ where
$$p_{j}(t) = m_{Thres} \left(\theta(t - (j-1)T) - \theta(t - jT)\right)$$
is a square input pulse (from a separate neuronal population) of length $T$, and $\tau$ is a synaptic timescale (for instance $3-11$ ms for AMPA neurons \citep{pmid8987749}). Here, the piecewise-linear activity function
$$f(m) = \left\{ \begin{array}{cc} 0,& m \le m_{Thres} \\ m - m_{Thres},& m > m_{Thres} \end{array} \right. \; ,$$
where $m_{Thres}$ is a firing threshold, is a good approximation to population simulation results \citep{Gerstner1995}, even for relatively strong synapses \citep{NordlieEtAl2010}.

For $0 < t < T$, population $j+1$ is excited by a gating pulse while receiving synaptic input from population $j$. With the ansatz $m_{j}(t) = Ae^{-t/\tau}$, $0 < t < T$, and assuming $A < m_{Thres}$, population $j+1$ obeys
$$\tau \frac{d}{dt}m_{j+1} = -m_{j+1} + Ae^{-t/\tau} \; , \; m_{j+1}(0) = 0$$ giving the solution $m_{j+1}(t) = SA\frac{t}{\tau} e^{-t/\tau}$.

For $T < t < \infty$, the gating pulse on population $j+1$ is turned off. Thus, population $j+1$ obeys $$\tau \frac{d}{dt}m_{j+1} = -m_{j+1} \; ,$$ with $m_{j+1}(T) = SA\frac{T}{\tau}e^{-T/\tau}$, giving the solution $m_{j+1}(t) = SA\frac{T}{\tau}e^{-T/\tau}e^{-(t-T)/\tau}$. For exact transfer, $m_{j+1}(t-T) = m_j(t)$, requiring that $S\frac{T}{\tau}e^{-T/\tau} = 1$. Thus, $S_{exact} = \frac{\tau}{T} e^{T/\tau}$.

In summary, we have the solution
\begin{eqnarray}
  m_{j+1}(t) = \left\{ \begin{array}{ll} SA \frac{t}{\tau}e^{-t/\tau}, & 0 < t < T \nonumber \\
                                                               SA \frac{T}{\tau}e^{-t/\tau}, & T < t < \infty \nonumber \end{array} \right.
\end{eqnarray}
and
\begin{eqnarray}
  p_{j+1}(t) = \left\{ \begin{array}{ll} m_{Thres}, & 0 < t < T \nonumber \\
                                                               0, & T < t < \infty \nonumber \end{array} \right.
\end{eqnarray}

In our previous mechanism, the circuit transferred synaptic current between an upstream and a downstream population. A pulse excited the upstream population into the firing regime and an ongoing inhibition kept the downstream population silent, except during the excitatory pulse. The solution was
$${I_{j+1}}\left( t \right) = \left\{ {\begin{array}{*{20}{ll}}
   {SA\frac{t}{\tau }{{\mathop{\rm e}\nolimits} ^{ - t/\tau }}}, & 0 < t < T  \\
   {SA\frac{T}{\tau }{{\mathop{\rm e}\nolimits} ^{ - t/\tau }}}, & T < t < \infty
\end{array}} \right.$$
and
$${m_{j+1}}\left( t \right) = \left\{ {\begin{array}{*{20}{ll}}
   0, & 0 < t < T  \\
  SA\frac{T}{\tau }{{\mathop{\rm e}\nolimits}^{ - t/\tau }}, & T < t < 2T \\ 
   0, & 2T < t < \infty
\end{array}} \right.$$
with
\begin{eqnarray}
  p_{j+1}(t) = \left\{ \begin{array}{ll} -I_0^{Inh}, & 0 < t < T \nonumber \\
                                                              I_0^{Exc}, & T < t < 2T \nonumber \\
                                                              -I_0^{Inh}, & 2T < t < \infty \nonumber\end{array} \right.
\end{eqnarray}
where $I_0^{Inh}$ is the amplitude of an ongoing inhibition, $I_0^{Exc}$ is the amplitude of an excitatory pulse. Note that the expressions for firing rate and current amplitude are the same, but that for the firing rate model the excitatory pulse acts during the increase in firing, while the excitatory pulse acts during the decay of the current. Note also, that the firing rate transfer model can be made to accommodate the transfer of larger firing rates ($A > m_{Thres}$) by the incorporation of inhibition into the model.

\subsection{Information Processing Using Exact Transfer Mechanisms}

Because either current or firing-rate amplitude transfer is in the linear regime, downstream computations may be considered as linear maps (matrix operations) on a vector of neuronal population amplitudes. For instance, consider a vector of neuronal populations with rates $\mathbf{m}^{j}$ connected via a connectivity matrix $K$ to $\mathbf{m}^{j+1}$:
$$\mathbf{m}^{j}(t) \overset{K}{\rightarrow} \mathbf{m}^{j+1}(t) \; .$$ 
With feedforward connectivity, given by the matrix $K$, the firing rate of population $i$ in layer $j+1$, $m^{j+1}_i$, obeys
$$\tau \frac{d}{dt} m^{j+1}_i = -m^{j+1}_i + f_i\left( \sum_k K_{ik} m^j_{k} + p_i(t) \right) \; .$$
This results in the solution $\mathbf{m}^{j+1}(t-T) = P K \mathbf{m}^j(t)$, where $P$ is a diagonal matrix with a vector, $\mathbf{p}$, of $0$s and $1$s on the diagonal indicating which neurons were pulsed during the transfer.

If the matrix of synaptic weights, $K$, were square and orthogonal, the transformation would represent an orthogonal change of basis in the vector space $\mathbb{R}^n$, where $n$ is the number of populations in the vector. Convergent and divergent connectivities would be represented by non-square matrices.

This type of information processing is distinct from concatenated linear maps in the sense that information may be dynamically routed via suitable gating. Thus, we can envision information manipulation by sets of non-abelian operators, i.e., with non-commuting matrix generators, that may be flexibly coupled. We can also envision introducing pulse-gated nonlinearities into our circuit to implement regulated feedback.

\subsection{Information Coding Framework}

Our discussion above has identified the three components of a unified framework for information coding: 
\begin{enumerate}
\item information content - graded firing rate, $\mathbf{m}$, or current, $\mathbf{I}$
\item information processing - synaptic weights, $K$
\item information control - pulses, $\mathbf{p}$
\end{enumerate}
Note that the pulsing control, $\mathbf{p}$, serves as a gating mechanism for routing neural information into (or out of) a processing circuit. We, therefore, refer to amplitude packets, $\mathbf{m}$ or $\mathbf{I}$, that are guided through a neural circuit by a set of stereotyped pulses as ``bound" information.

Consider one population coupled to multiple downstream populations. Separate downstream processing circuits may be multiplexed by pulsing one of the set of downstream circuits. Similarly, copying circuit output to two (or more) distinct downstream populations may be performed by pulsing two populations that are identically coupled to one upstream population.

In order to make decisions, non-linear logic circuits are required. Many of these are available in the literature \citep{6707077,VogelsAbbott2005}. Simple logic gates are straightforward to construct within our framework by allowing interaction between information control and content circuits. For instance, to construct an AND gate, use gating pulses to feed two sub-threshold outputs into a third population, if the inputs are $(0,0)$, $(0,1)$ or $(1,0)$, none of the combined pulses exceed threshold and no output is produced. However, the input $(1,1)$ gives rise to an output pulse. Other logic gates, including the NOT may be constructed, giving a Turing complete set of logic gates. Thus, these logic elements may be used for plastic control of functional connectivity, i.e. the potential for rapidly turning circuit elements on or off, enabling information to be dynamically routed.

\subsection{A High-Fidelity Memory Circuit}

\begin{figure}[]
  \includegraphics[width=\columnwidth]{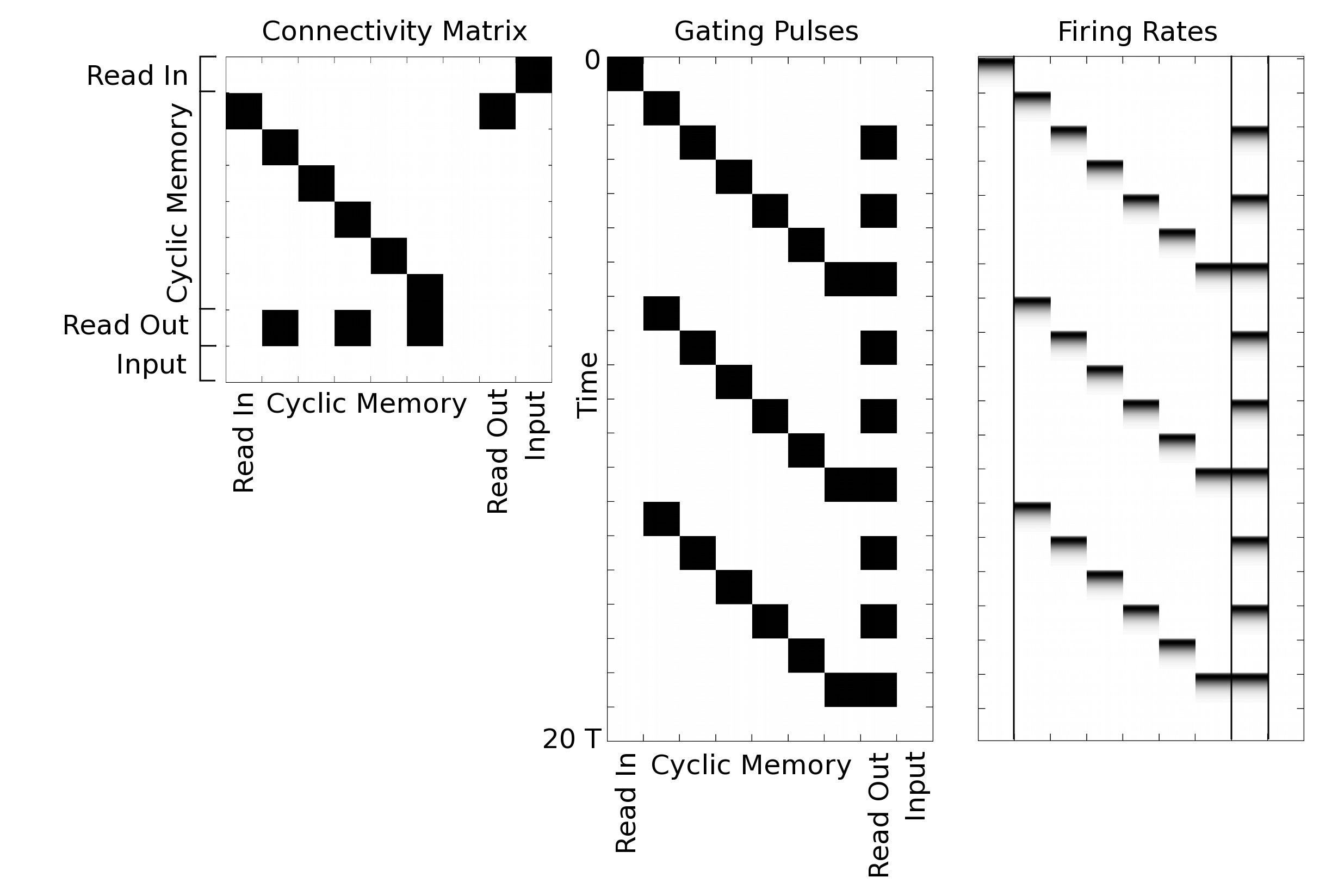}
  \caption{Memory circuit maintaining a single graded firing rate. (Left) Connectivity Matrix. White denotes $0$ entries and black denotes $1$s. The connectivity matrix is subdivided into four sets of rows. ``Input" designates filtered input from an outside source. The first row connects the ``Read In'' population to the input. The Read In population transduces the filtered input into a graded firing rate packet that then propagates through the memory circuit. The ``Cyclic Memory" contains cyclically connected, feedforward populations around which the graded packet is propagated. The ``Read Out" population is postsynaptic to every other population in the Cyclic Memory and may be used to transfer graded packets at high frequencies to another circuit. (Middle) Gating Pulses. White denotes $0$, black denotes $m_{Thres}$. $T/\tau = 8$. This sequence of gating pulses is used to bind and propagate the graded memory. Time runs from top to bottom. We show three complete cycles of propagation. The initial pulse on the Read In population binds the filtered input. The subsequent pulses within the cyclic memory rotate the packet through the memory populations. The pulses in the Read Out population copy the memory to a distinct population, which could be in another circuit. (Right) Firing Rates. White denotes $0$, black denotes the maximum for this particular firing rate packet. The input (seen only faintly in the right hand column) is transduced into the Read In population after time $0$ (upper left of panel). The memory is subsequently propagated through the circuit and copied from every other population to the Read Out.}
\end{figure}

As a first complete example, we demonstrate a memory circuit. Our circuit generalizes the IM model by allowing for graded memory and arbitrary multiplexing of memory to other neural circuits. Because it is a population model, it is more robust to perturbations than the IM model, which transfers spikes between individual neurons. It is different from the IM model in that our circuit retains only one graded amplitude, not many (although this could be arranged) \citep{LismanIdiart1995}. However, our model retains the multiple timescales that generate theta and gamma oscillations from pulse gating inherent to the IM model \citep{LismanIdiart1995}. Additionally, other graded memory models \citep{SeungEtAl2000,Goldman2008} make use of relatively large time constants that are larger even than NMDA timescales, whereas ours makes use of an arbitrary synaptic timescale, $\tau$, which may be modified to make use of any natural timescale in the underlying neuronal populations, including AMPA or NMDA. Our model is based on exact, analytical expressions, and because of this, the memory is infinitely long-lived at the mean-field level (until finite-size effects and other sources of variability are taken into account).

The circuit has four components, a population for binding a graded amplitude into the circuit (`read in'), a cyclical memory, a `read out' population meant to emulate the transfer of the graded amplitude to another circuit, and an input population. The memory is a set of $n$ populations coupled one to the other in a circular chain with one of the populations (population $1$) receiving gated input from the read in population. Memory populations $j$ and $j+1$ receive coherent, phase shifted (by time $jT$) pulses that transfer the amplitude around the chain. In this circuit, $n$ must be large enough that when population $n$ transfers its amplitude back to population $1$, population $1$'s amplitude has relaxed back to (approximately) zero. The read out is a single population identically coupled to every other population in the circular chain. This population is repeatedly pulsed allowing the graded amplitude in the circular chain to be repeatedly read out.

In Fig. 1, we show an example of the memory circuit described here with $n = 6$. The gating pulses sequentially propagate the graded firing rate amplitude around the circuit. The read out population is coupled to every other population in the memory. Thus, in this example, the oscillation frequency of the read out population is three times that of the memory populations, i.e. theta-band frequencies in the memory populations would give rise to gamma-band frequencies in the read out.

This memory circuit, and other circuits that we present below, has the property that the binding of information is instantiated by the pulse sequence and is independent of the information carried in graded amplitudes and also independent of synaptic processing. Because of the independence of the control apparatus from information content and processing, this neural circuit is an automatic processing pathway whose functional connectivity (both internal and input/output) may be rapidly switched on or off and coupled to or decoupled from other circuits. We propose that such dynamically routable circuits, including both processing and control components, are the neural correlates of automatic cognitive processes that have been termed zombie modes \citep{CrickKoch2003}.

\subsection{A Moving Window Fourier Transform}

\begin{figure}[]
  \includegraphics[width=\columnwidth]{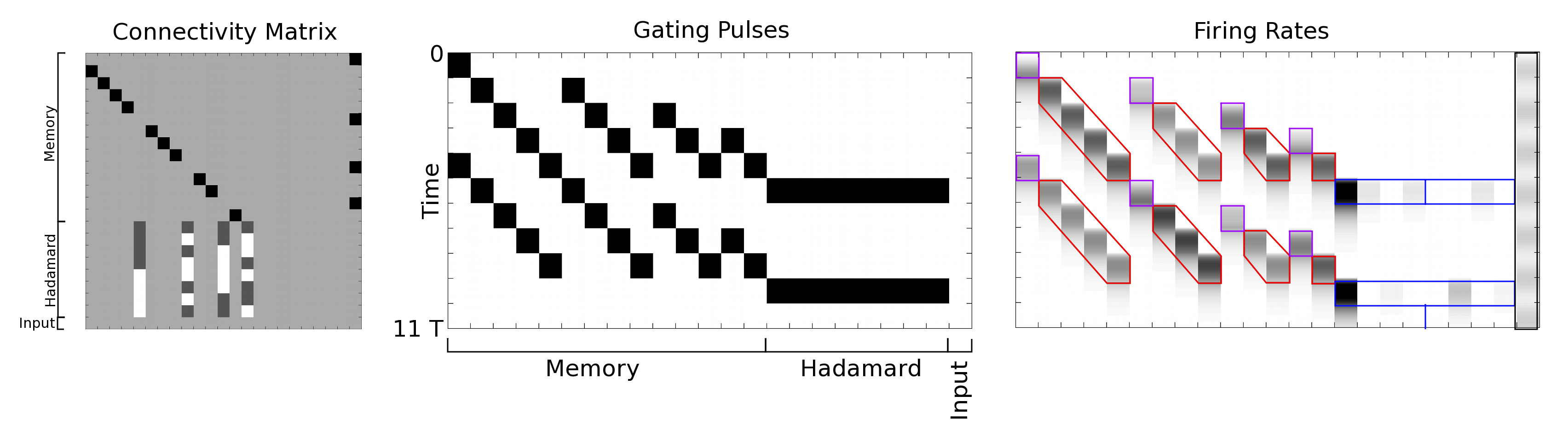}
  \caption{$4 \times 4$ Hadamard transform on a window of input values moving in time. (Left) Connectivity Matrix. White denotes $-1/2$, light gray denotes $0$, dark gray denotes $1/2$, and black denotes $1$. The connectivity matrix is subdivided into three sets of rows. ``Memory" designates Read In and (non-cyclic) Memory populations. ``Hadamard" designates populations for the calculation of Hadamard coefficients. Because the packet amplitudes can only be positive, the Hadamard transform is divided into two parallel operations, one that results in positive coefficients and one that results in absolute values of negative coefficients. ``Input" designates filtered input from an outside source. (Middle) Gating Pulses. White denotes $0$, black denotes $m_{Thres}$. $T/\tau = 2$. Time runs from top to bottom. We show the computation for two successive windows, each of length $4T$. The pulses transduce the input into four memory chains of length $4T$, $3T$, $2T$ and $T$. Thus, four temporally sequential inputs are bound in four of the memory populations beginning at times $t = 4, 8 T$. Hadamard transforms are performed beginning at $t = 5, 9 T$. Note that the second read in starts one packet length before the Hadamard transform so that the temporal windows are adjacent. (Right) Firing Rates. White denotes $0$, black denotes the maximum firing rate. Purple outlines denote Read In, red denote Memory and blue denote Hadamard transform populations. The left four Hadamard outputs are positive coefficients. The right four are absolute values of negative coefficients. The sinusoidal input waveform is shown to the right.}
\end{figure}

The memory circuit above used one-to-one coupling. It was simple in that information was copied, but not processed. Our second example demonstrates how more complex information processing may be accomplished within a zombie mode. With a simple circuit that performs a Hadamard transform (a Fourier transform using square-wave-shaped Walsh functions as a basis), we show how streaming information may be bound into a memory, then processed via synaptic couplings between populations in the circuit.

A set of read in populations are synaptically coupled to the input. A set of memory chains are coupled to the read in. The final population in each memory chain is coupled via a connectivity matrix that implements a Hadamard transform. Gating pulses cause successive read in and storing in memory of the input, until the Hadamard transform is performed once the memory contains all successive inputs in a given time window simultaneously. Because the output of the Hadamard transform may be negative, two populations of Hadamard outputs are implemented, one containing positive coefficients, and another containing absolute values of negative coefficients.

In Fig. 2, we show a zombie mode where four samples are bound into the circuit from an input, which changes continuously in time. Memory populations hold the first sample over four transfers, the second sample over three transfers, etc. Once all samples have been bound within the circuit, the Hadamard transform is performed with a pulse on the entire set of Hadamard read out populations. While this process is occurring, a second sweep of the algorithm begins and a second Hadamard transform is computed.

The connectivity matrix for the positive coefficients of the Hadamard transform was given by
\begin{equation} \nonumber
   H = \frac{1}{2}\left[ \begin{array}{cccc} 1 & 1 & 1 & 1 \\ 1 & -1 & 1 & -1 \\ 1 & 1 & -1 & -1 \\ 1 & -1 & -1 & 1 \end{array} \right] \; ,
\end{equation}
and the absolute values of the negative coefficients used the transform $-H$.

\subsection{A Re-entrant Spatial Rotation Circuit}

\begin{figure}[]
  \includegraphics[width=\columnwidth]{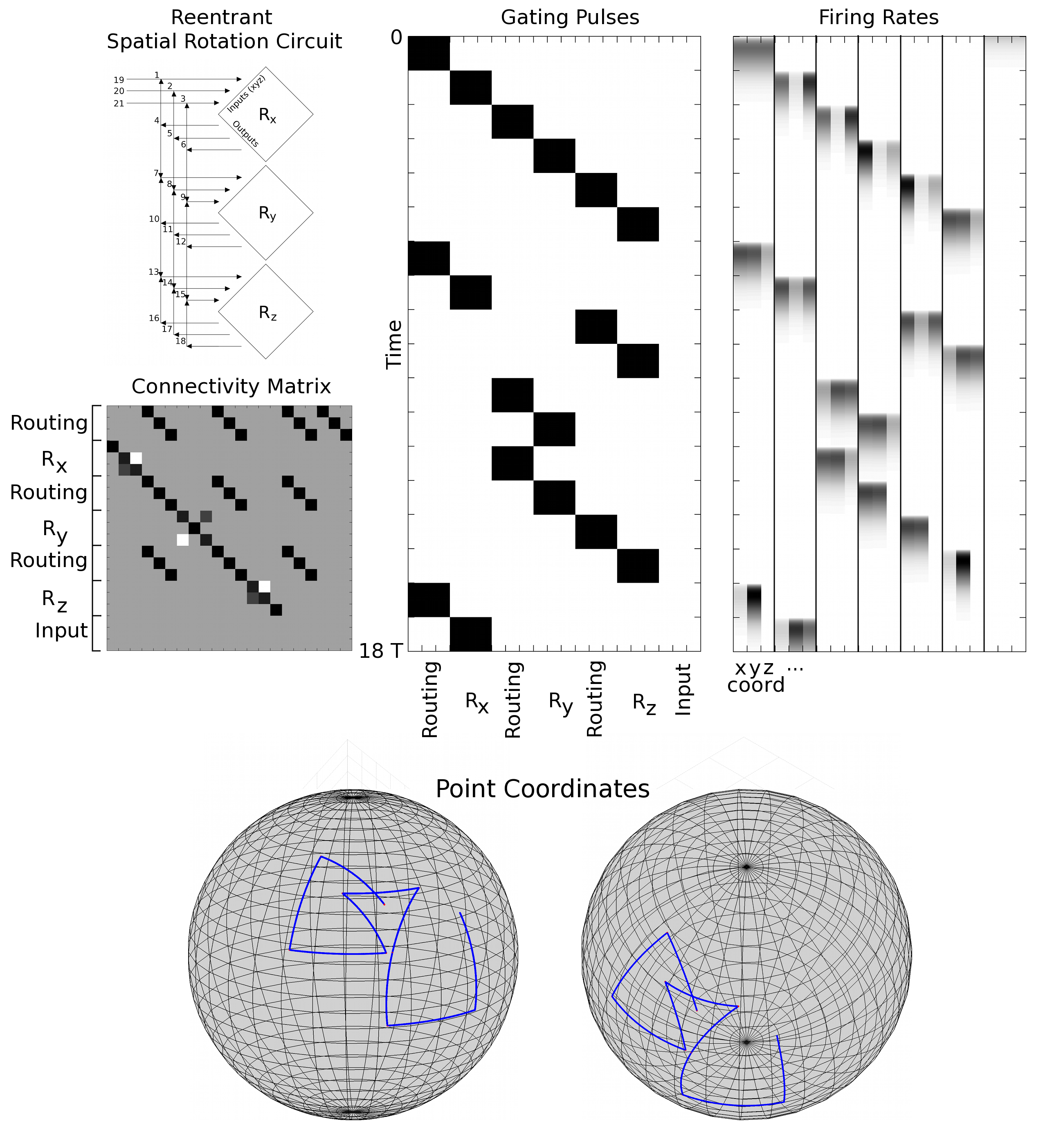}
  \caption{Spatial rotation of three-dimensional coordinates. (Left, Top) A diagram of the circuit. Diamonds represent spatial rotation about the given axis, $x$, $y$, or $z$, by angle $2\pi/10$. $R_x$ takes coordinates $1$, $2$ and $3$ as input and outputs $4$, $5$ and $6$. $R_y$ takes $7$, $8$ and $9$ as inputs and outputs $10$, $11$ and $12$. And $R_z$ takes $13$, $14$ and $15$ as input and outputs $16$, $17$ and $18$. Outputs may be routed to any of $R_x$, $R_y$ or $R_z$ giving a reentrant circuit. (Left, Bottom) Connectivity Matrix. Light gray denotes $-\sin{(2\pi/10)}$, gray denotes $0$, dark gray denotes $\sin{(2\pi/10)}$, black denotes $1$, with some black squares representing $\cos{(2\pi/10)}$ indistinguishable from $1$. Input may be read into $1$, $2$ and $3$ only. Outputs of the rotations may be read into any of $(1,2,3)$, $(7,8,9)$ or $(13,14,15)$ and subsequent rotations performed on these amplitudes. (Middle) Gating Pulses. White denotes $0$, black denotes $m_{Thres}$. $T/\tau = 3$. Time runs from top to bottom.  We show read in, then routing of the coordinates through $R_x$, $R_y$, $R_z$, $R_x$, $R_z$, $R_y$, $R_y$, $R_z$, and $R_x$, successively. (Right) Firing Rates. Initially uniform coordinates $(x, y, z) = (1, 1, 1)$, are successively rotated about the various axes. (Bottom) Two views of the coordinates connected by geodesics as they are rotated.}
\end{figure}

Our final example makes use of plastic internal connectivity to perform an arbitrary set of rotations of a vector on the sphere. Three fixed angle rotations about the $x$, $y$  and $z$ axes are arranged such that the output from each rotation may be copied to the input of any of the rotations. Because the destination is determined by the pattern of gating pulses, this circuit is more general than a true zombie mode because it is not automatic: manipulation of the rotations would be expected to occur from a separate routing control circuit.

In Fig. 3, initial spatial coordinates of $(1, 1, 1)$ were input to the circuit. The pulse sequence rotated the input first about the $x$-axis, then sequentially about $y$, $z$, $x$, $z$, $y$, $y$, $z$, $x$ axes. Views from two angles illustrate the rotations that were performed by the circuit.

This circuit demonstrates the flexibility of the information coding network that we have introduced. It shows a complex circuit capable of rapid computation with dynamic routing, but with a fixed connectivity matrix. Additionally, it is an example of how a set of non-commuting generators may be used to form elements of a non-abelian group within our framework.

\section{Discussion}

The discovery of exact firing-rate and current transfer mechanisms has allowed us to construct a conceptual framework for the active manipulation of information in neural circuits. The most important aspect of this type of information coding is that it separates control of the flow of information from information processing and the information itself.

The four functions that must be served by a neural code are \citep{PerkelBullock1968,pmid20725095}: stimulus representation, interpretation, transformation and transmission. Our framework serves all of these functions. From last to first: the exact transfer mechanisms that we have identified serve the transmission function; synaptic couplings provide the capability of transforming information; the pulse dependent, selective read out of information serves the interpretation function; and read in populations, as we demonstrated in our examples, may be used to convert stimulus information into a bound information representation.

The current and firing rate transfer mechanisms are sufficiently flexible that the pulses used for gating may be of different durations depending on the pulse length, $T$, and the time constant, $\tau$, of the neuronal population involved. Thus, the mechanism could be used either for information coding in populations (as demonstrated above) or for labeled line coding using temporal rate averaging.

The separation of control populations from those representing information content distinguishes our framework from mechanisms such as CTC, where communication between neuronal populations depends on the coincidence of integration windows in phase-coherent oscillations. In the CTC mechanism, information containing spikes must coincide to be propagated. In our framework, information containing spikes must coincide with gating pulses that enable communication. In this sense, it is `communication through coherence with a control mechanism'.

The separation of control and processing has further implications, one of which is that, as noted above, while a given zombie mode is processing incoming information, one does not expect the pulse sequence to change dependent on the information content. This has been seen in experiment, and presented as an argument against CTC in the visual cortex \citep{pmid12540900}, but is consistent with our framework.

The basic unit of computation in our framework is a pulse gated transfer. Given this, we suggest that each individual pulse within an oscillatory set of pulses represents the transfer and processing of a discrete information packet. For example, in a sensory circuit that needs to quickly and repeatedly process a streaming external stimulus, short pulses could be repeated in a stereotyped, oscillatory manner using high-frequency gamma oscillations to rapidly move bound sensory information through the processing pathway. Whereas circuits that are used occasionally or asynchronously might not involve oscillations at all, just a precise sequence of pulses that gate a specific information set through a circuit. A possible example of this is bat echo-location, an active sensing mechanism, where coherent oscillations have not been seen \citep{pmid22051680}.

Based on our coding framework, we can envision an architecture for the brain that is divided into separate components for 1) directing the flow of information, 2) processing bound information, and 3) regulating and maintaining optimal information transfer. Because control is enacted via precisely timed pulses, neural pattern generators would be expected to be information control centers. Cortical pattern generators, such as those proposed by Yuste \citep{YusteEtAl2005}, because of their proximity to cortical circuits, could logically be the substrate for zombie mode control pulses. They would be expected to generate sequential, stereotyped pulses to dynamically route information flow through a neural circuit, as has been found in rat somatosensory cortex \citep{pmid17185420}. Global routing of information via attentional processes, on the other hand, would be expected to be performed from brain regions with broad access to the cortex, such as the thalamus.

Recent evidence shows that parvalbumin-positive basket cells (PVBCs) can gate the conversion of current to spikes in the amygdala \citep{pmid24814341}. Also,  PVBCs and oriens lacunosum moleculare (OLM) cells have been implicated in precision spiking related to gamma- and theta-oscillations \citep{pmid23010933} and shown to be involved in memory-related structural-plasticity \citep{pmid12594513}. Therefore, zombie mode pattern generators would likely be based on a substrate of these neuron types.

\begin{acknowledgments}
{\bf Acknowledgements} L.T. thanks the UC Davis Mathematics Department for its hospitality. This work was supported by the Ministry of Science and Technology of China through the Basic Research Program (973) 2011CB809105 (L.T.) and the Natural Science Foundation of China grant 91232715 (L.T.).
\end{acknowledgments}

\bibliographystyle{unsrtnat}
\bibliography{Biblio}

\end{document}